\documentclass[crop]{CSL}
\jname{International Journal of Astrobiology}

\usepackage{url}
\usepackage{multirow}
\usepackage{graphicx}
\usepackage{epstopdf}
\usepackage{nameref}
\usepackage{hyperref}
\hypersetup{colorlinks=true, citecolor = blue, urlcolor = blue, linkcolor = blue}

\begin{document}

\supertitle{Research Paper (Preprint)}

\title[{\tt Earth-Like}: exploring planet diversity]{{\tt Earth-Like}: An education \& outreach tool for exploring the diversity of planets like our own}

\author[E. J. Tasker {\emph et al.}]{Elizabeth J. Tasker$^{1}$, Kana Ishimaru$^{2, 3}$, Nicholas Guttenberg$^{4}$ and Julien Foriel$^{4, 5}$}

\corres{\name{Elizabeth J. Tasker} \email{elizabeth.tasker@jaxa.jp}}

\address{\auadd{1}{Institute of Space and Astronautical Sciences, Japan Aerospace Exploration Agency, Yoshinodai 3-1-1, Sagamihara, Kanagawa 252-5210, Japan}; \auadd{2}{Department of Earth \& Planetary Environmental Science, The University of Tokyo, 7-3-1 Hongo, Tokyo 113-0033, Japan}; \auadd{3}{Lunar and Planetary Laboratory, University of Arizona, Tucson, AZ, USA}; \auadd{4}{Earth-Life Science Institute, Tokyo Institute of Technology, 2-12-1 Tokyo, Japan} and \auadd{5}{Harvard University, Department of Earth and Planetary Sciences, Cambridge, MA 02138, USA}}

\begin{abstract}
{\tt Earth-Like} is an interactive website and twitter bot that allows users to explore changes in the average global surface temperature of an Earth-like planet due to variations in the surface oceans and emerged land coverage, rate of volcanism (degassing), and the level of the received solar radiation. The temperature is calculated using a simple carbon-silicate cycle model to change the level of $\rm CO_2$ in the atmosphere based on the chosen parameters. The model can achieve a temperature range exceeding $-100^\circ$C to $100^\circ$C by varying all three parameters, including freeze-thaw cycles for a planet with our present-day volcanism rate and emerged land fraction situated at the outer edge of the habitable zone. To increase engagement, the planet is visualised by using a neural network to render an animated globe, based on the calculated average surface temperature and chosen values for land fraction and volcanism. The website and bot can be found at {\tt earthlike.world} and on twitter as {\tt @earthlikeworld}. Initial feedback via a user survey suggested that {\tt Earth-Like} is effective at demonstrating that minor changes in planetary properties can strongly impact the surface environment. The goal of the project is to increase understanding of the challenges we face in finding another habitable planet due to the likely diversity of conditions on rocky worlds within our Galaxy.
\end{abstract}



\accepted{2019-12-05}

\maketitle

\section{Introduction}

In the last 30 years, we have gone from knowing only the planets of our own Solar System to discovering thousands of worlds orbiting other stars. Approximately two-thirds of these discoveries have been planets with radii less than twice that of the Earth\footnote[1]{NASA Exoplanet Archive \url{https://exoplanetarchive.ipac.caltech.edu/}}, leading both the general public and scientific community alike to ask the question: could any of these planets be habitable?

\parindent=0.5cm One of the challenges with addressing this question is communicating the probable diversity of planet environments. For example, changes in the distribution of sunlight on Earth driven by the small orbital adjustments of the Milankovitch cycles have triggered periods of global glaciation on roughly forty to hundred thousand year cycles \citep[e.g.][]{clark2006, hays1976}. Such variations in conditions are tiny compared with other planetary systems where changes in properties such as planet size, composition, orbit and stellar type are all up for grabs. The range of possible surface environments on these newly discovered worlds is therefore vast and understanding which conditions might be habitable is a major focus for planetary science and astrobiology in the coming decades.

However, present observations cannot directly demonstrate diversity in surface properties. Our current knowledge of individual exoplanets is typically restricted to a measurement of the planet's bulk size (either radius or minimum mass depending on the detection technique) and the level of radiation received from the star. This makes it impossible to measure surface conditions, or comment quantitatively on the likelihood a particular planet could support life \citep{tasker2017}.

Communicating both this potential diversity and present observational restrictions beyond the exoplanet community has been difficult. Exoplanets with a size consistent with a rocky surface and whose orbit sits within the so-called habitable zone are frequently portrayed in the main-stream media and even in scientific press releases as having a high probability of supporting similar surface conditions to Earth, with terms such as `Earth's cousin' or `Earth 2.0' being in common use. This leads to the impression that rocky planet diversity is minimal

\Fpagebreak

\noindent and habitability is simply a function of planet size and radiation levels. Pervasion of these views are not only a failure to communicate scientific results, but risk the credibility of the field and the support for future instruments capable of probing surface conditions on smaller worlds due to such information being deemed as already available.

While we cannot yet observe the diversity of rocky exoplanets, we can use models to explore potential surface environments \citep[e.g.][]{DelGenio2019, Unterborn2018, Heng2011, Pierrehumbert2011}. Simulations are an excellent way to investigate the impact of variations in different properties, allowing a broad range of systems to be compared. However, techniques such as global climate models are computationally intensive and require technical knowledge to run and analyse the results. This makes them out of reach for use as teaching and outreach tools or even for use in the discussion section of papers whose primarily focus is observational or otherwise not simulation based.

In this paper, we present {\tt Earth-Like} as an interactive tool to explore a subset of the diversity in rocky planets through variations in surface ocean and exposed land fraction, volcanism (degassing) and location within the habitable zone from the present-day Earth. Values for the three parameters can be selected via a simple web-based interface and the model run online. These choices are all for properties that have varied during the Earth's own history and are poorly constrained in that there is no reason to expect the values to remain the same even for a planet forming within an exact Solar System analogue. {\tt Earth-Like} can be used to qualitatively understand the diversity in terrestrial planets that have a liquid body of water and sit within the habitable zone, and point to the probable diversity of worlds whose properties are far more varied.

{\tt Earth-Like} calculates the surface temperature of the planet based on how the chosen parameter values impact a simple model of the carbon-silicate cycle. Both the results and information on the model are presented on the website through multiple channels, including video, text and images. This is designed to increase the accessibility of the site to all users from school students to the interested general public. Section `\nameref{sec:climate}' in this paper describes the climate model used in {\tt Earth-Like} and the choice and range in available parameters. Visualising the planet and the neural network will be discussed in section `\nameref{sec:visual}'. Section `\nameref{sec:interface}' looks at the interface for the model and section `\nameref{sec:feedback}' details the feedback from the site questionnaire. We summarise the project in the `\nameref{sec:conclusions}' section.

\section{The climate model}
\label{sec:climate}

{\tt Earth-Like} calculates a global surface temperature by using a box model of the carbon-silicate cycle to find the abundance of carbon in the atmosphere. The carbon-silicate cycle is one of the most important processes on Earth for maintaining temperate conditions. It is a long-term geochemical cycle that regulates the concentration of atmospheric $\rm CO_2$ by circulating carbon between the atmosphere, ocean, seafloor and mantle over time scales exceeding $10^5$ years. As $\rm CO_2$ is a greenhouse gas that absorbs radiation, changes to the concentration of this molecule in the atmosphere affect the planet's global climate.

During the carbon-silicate cycle, $\rm CO_2$ is drawn out of the atmosphere through weathering with silicate rocks and transferred in rivers and ground water to the ocean. Carbonate minerals form in aqueous solution and are deposited in sedimentary rocks. Subduction takes these from the seafloor where high temperatures result in the carbon being released and returned to the atmosphere via volcanism and degassing where it reforms $\rm CO_2$.

The silicate weathering that draws the carbon from the atmosphere is a negative feedback process that operates faster in higher surface temperatures. When conditions are warmer, weathering draws $\rm CO_2$ more rapidly from the atmosphere which allows radiation to more easily escape into space and cools the planet. Conversely, a cooler environment slows weathering to build-up $\rm CO_2$ in the atmosphere and trap radiation more efficiently to keep the planet from freezing \citep{Walker1981}. This feedback process is thought to have been an important factor in maintaining liquid surface water on the Earth during the first 2 billion years of our planet's history when the incident flux from the young Sun was substantially weaker than today \citep{Feulner2012}.

The regulation of the surface temperature through adjustments in the level of atmospheric $\rm CO_2$ by the carbon-silicate cycle define the boundaries of the circumstellar classical habitable zone. The classical habitable zone is a region around a star where the Earth's carbon-silicate cycle can keep the surface temperature suitable for liquid water for the range of incident flux \citep{hz, Kasting1993}. Orbiting within the habitable zone does not provide any guarantee of a planetary environment capable of supporting liquid water, but a planet with the same surface pressure and geochemical processes as the Earth could maintain surface oceans and would therefore be most likely to support life in this region \citep{Seager2013}. For that reason, the habitable zone is used a target selection tool for identifying planets for astrobiological follow-up studies.

\begin{figure}
    \centerline{\includegraphics[width=0.9\columnwidth]{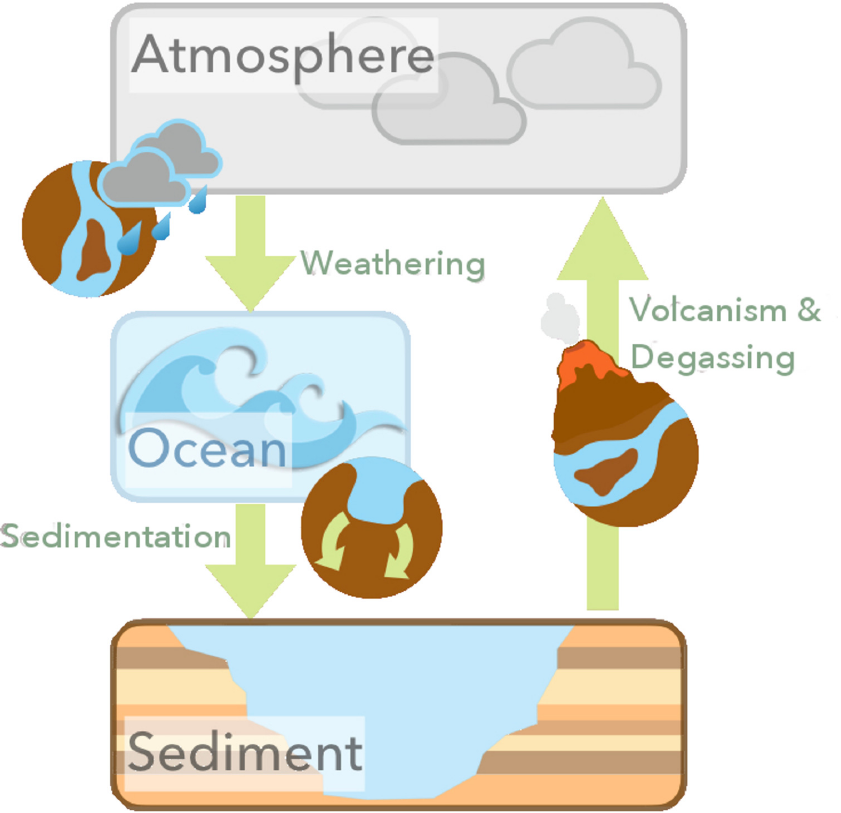}}
    \caption{Schematic illustration of the three-box carbon-silicate cycle model used by {\tt Earth-Like}. Carbon is exchanged between three reservoirs in the atmosphere, ocean and sediment (ocean floor and mantle) through fluxes for weathering, $F_w$, sedimentation, $F_{\rm sed}$, and degassing, $F_{\rm vol}$.}
    \label{fig:boxmodel}
\end{figure}

{\tt Earth-Like} models the carbon-silicate cycle using mass balance equations for circulating carbon between three reservoirs: the atmosphere ($R_{\rm atm}$), ocean ($R_{\rm oce}$) and sediment (ocean crust and mantle) ($R_{\rm sed}$). The flux of carbon from the atmosphere to ocean is the silicate weathering flux, $F_{\rm w}$, the flux from the ocean to sediment is the sedimentation or subduction flux, $F_{\rm sed}$, and the flux from the sediment back to the atmosphere is the rate of volcanism and degassing, $F_{\rm vol}$. This is shown schematically in Figure~\ref{fig:boxmodel}. The carbon mass in each reservoir is then given by integrating over time:
\begin{align*}
\frac{dR_{\rm atm}}{dt}     &=  F_{\rm vol} - F_{\rm w} \\
\frac{dR_{\rm oce}}{dt}     &=  F_{\rm w} - F_{\rm sed} \\
\frac{dR_{\rm sed}}{dt}     &=  F_{\rm sed} - F_{\rm vol} \\
\frac{dR_{\rm atm}}{dt} &+ \frac{dR_{\rm oce}}{dt} + \frac{dR_{\rm sed}}{dt} = 0
\end{align*}
\noindent We assume that the silicate weathering flux depends on the fraction of the planet surface that is covered by land (the weathering area), $\gamma$, the abundance of $\rm CO_2$ in the atmosphere and the surface temperature, $T_s$. This takes the form \citep{Abbot2012, Foley2015, Walker1981}:
\begin{equation}
F_{\rm w} = F_{\oplus}\left(\frac{\gamma}{\gamma_\oplus}\right)\left(\frac{R_{\rm atm}}{R_{\rm atm,0}}\right)^\beta e^{(T_s - T_{s,\oplus})/k}
\end{equation}
\noindent where the parameters with the $\oplus$ subscript denote that value on present-day Earth (see Table~\ref{table:symbols}). The index $\beta = 0.165$ is a rate constant that controls the dependence of weathering on $\rm CO_2$ abundance and can take values between 0 and 1 \citep{Berner1994, Abbot2012, krissansen2017}. Our value is selected based on the behaviour at the inner and outer edges of the habitable zone (see section~\nameref{sec:parameters}). $k = 10$\,K controls the response of weathering to surface temperature, with a value consistent with analysis from weathering data \citep{West2005}. Although weathering can occur on the seafloor if the ocean depth is not high enough for the formation of high pressure ices, this rate is reduced compared to weathering on exposed land and not directly linked to surface temperature. We therefore ignore this for simplicity in this work.

\begin{table}
\tabcolsep4pt
\processtable{Constants used in the climate model, based on recent Earth values.\label{table:symbols}}{
\begin{tabular}{p{1.5cm} p{4cm} p{2.5cm}}
\hline
\rowcolor{Theadcolor} Symbol & Definition & Baseline Value\\\hline
R$_{\rm atm, \oplus}$ & Atmosphere carbon reservoir & 600.0 Gtns$^\dagger$  \\
R$_{\rm oce, \oplus}$ & Ocean carbon reservoir & $3.8 \times 10^4$\,Gtns$^\ddagger$  \\
R$_{\rm sed, \oplus}$ & Sediment carbon reservoir & $4.8\times 10^7$\,Gtns$^{\dagger\dagger}$   \\
F$_{\oplus}$ & Net flux of ${\rm CO_2}$ from atmosphere to ocean reservoirs. & $7\times 10^4$\,Gtns/Myr\,$^\mathsection$  \\
T$_{\rm s, \oplus}$ & Global surface temperature & 288\,K   \\
T$_{\rm eq, \oplus}$ & Equilibrium temperature & 255\,K   \\
$\gamma_\oplus$ & Exposed land fraction & 0.29  \\
$\alpha_\oplus$ & Surface albedo & 0.3\\
\hline
\end{tabular}}{
\begin{tablenotes}
\item $^\dagger$\,Pre-industrial value, IPCC assessment \citep{ipccatm}, $\ddagger$\,IPCC assessment \citep{ipccocean}, $\dagger\dagger$\, \cite{Caldeira1991}.
\item $\mathsection$\,The flux from silicate weathering on present day Earth is estimated at $1.4\times 10^5$\,GtC/Myr, but half of this is re-released to the atmosphere during carbonate formation. The net flux of ${\rm CO_2}$ from atmosphere to ocean reservoirs is therefore half the initial weathering flux \citep{Berner1994, Gaillardet1999, Foley2015}.
\end{tablenotes}}
\end{table}

The surface temperature is related to the abundance of $\rm CO_2$ in the atmosphere via the parameterisation developed in \citet{Walker1981}:
\begin{equation}
T_{\rm s} = T_{\rm s, \oplus} + 2(T_{\rm eq} - T_{\rm eq, \oplus}) + 4.6\left(\frac{R_{\rm atm}}{R_{\rm atm, \oplus}}\right)^{0.364} - 4.6
\label{eq:temp}
\end{equation}
\noindent where $T_{\rm eq}$ is the equilibrium temperature of the planet given by:
\begin{equation}
T_{\rm eq} = (1 - \alpha)\left(\frac{L_\odot}{16\pi\sigma a^2}\right)^{1/4}
\end{equation}
\noindent for solar luminosity, $L_\odot = 3.828 \times 10^{26}$\,W, Stefan-Boltzmann constant, $\sigma = 5.67\times 10^{-8}$\,Wm$^{-2}$K$^{-4}$ and the distance of the planet from the star, $a$. We use the following simple parameterisation for the albedo, $\alpha$, dependence on the surface temperature:
\begin{equation}
\alpha = \begin{cases}
     \alpha_\oplus & \text{$T_s > 273.15$\,K} \\
     \alpha_\oplus+(0.6-\alpha_\oplus)\left(\frac{273.15 - T_s}{273.15 - 250.0}\right) & \text{$273.15 > T_s > 250.0$\,K} \\
    0.6 & \text{$T_s \le 250.0$\,K}
    \end{cases}
\label{eq:albedo}
\end{equation}
\noindent Warmer planets have a surface albedo consistent with the present day Earth value of $\alpha_\oplus = 0.3$. Colder planets with a larger fraction of their surface covered with reflective ice have a higher albedo that reaches 0.6 for a frozen world, similar to that of the Jovian icy moon, Europa.

The degassing and the sedimentation fluxes are modelled as simple draining flows that are respectively proportional to the abundance of carbon in sediment and oceans, multiplied by the rate at which carbon is leaving the atmosphere:
\begin{align*}
F_{\rm vol} &= \Gamma F_{\oplus}\frac{R_{\rm sed}}{R_{\rm sed,\oplus}} \\
F_{\rm sed} &= F_{\oplus}\frac{R_{\rm oce}}{R_{\rm oce,\oplus}}
\end{align*}
\noindent The rate of degassing can be increased or slowed by a factor of $\Gamma$. This value is a parameter that can be selected by the user (see section `\nameref{sec:parameters}').

\subsection{Climate control parameters}
\label{sec:parameters}

\begin{table}
\tabcolsep4pt
\processtable{Climate parameters.\label{table:parameters}}{
\begin{tabular}{p{1.2cm} p{1.6cm} p{3.2cm} p{1.8cm}}
\toprule
\rowcolor{Theadcolor} Toolkit & Parameter & Description & Accepted range \\
\hline
\multirow{3}{*}{\it Classic} & $\gamma$ & Exposed land fraction & $0 - 1$ \\
&$\Gamma$ & Degassing factor & $0 - 1000$  \\
& HZ & Position within the habitable zone & $0 - 1$   \\
\hline
\multirow{4}{*}{\it Advanced} &  $\gamma$ & Exposed land fraction & $0 - 1$ \\
& $\Gamma$ & Degassing factor & $0 - 1000$ \\
& Stellar type & Luminosity of star & 0.0005 - 5.0\,L$_\odot^\dagger$\\
& a & Distance from the star (au) & $> 0$ \\
\end{tabular}}{
\begin{tablenotes}
\item $\dagger$ Stellar type can be selected from a pull-down menu with options ultracool ($0.0005$\,$L_\odot$), red dwarf ($0.002$\,$L_\odot$), orange dwarf ($0.5$\,$L_\odot$), yellow dwarf ($1.0$\,$L_\odot$) and yellow-white dwarf ($5.0$\,$L_\odot$).
\end{tablenotes}}
\end{table}

\begin{figure*}[!t]
    \centerline{\includegraphics[width=0.9\textwidth]{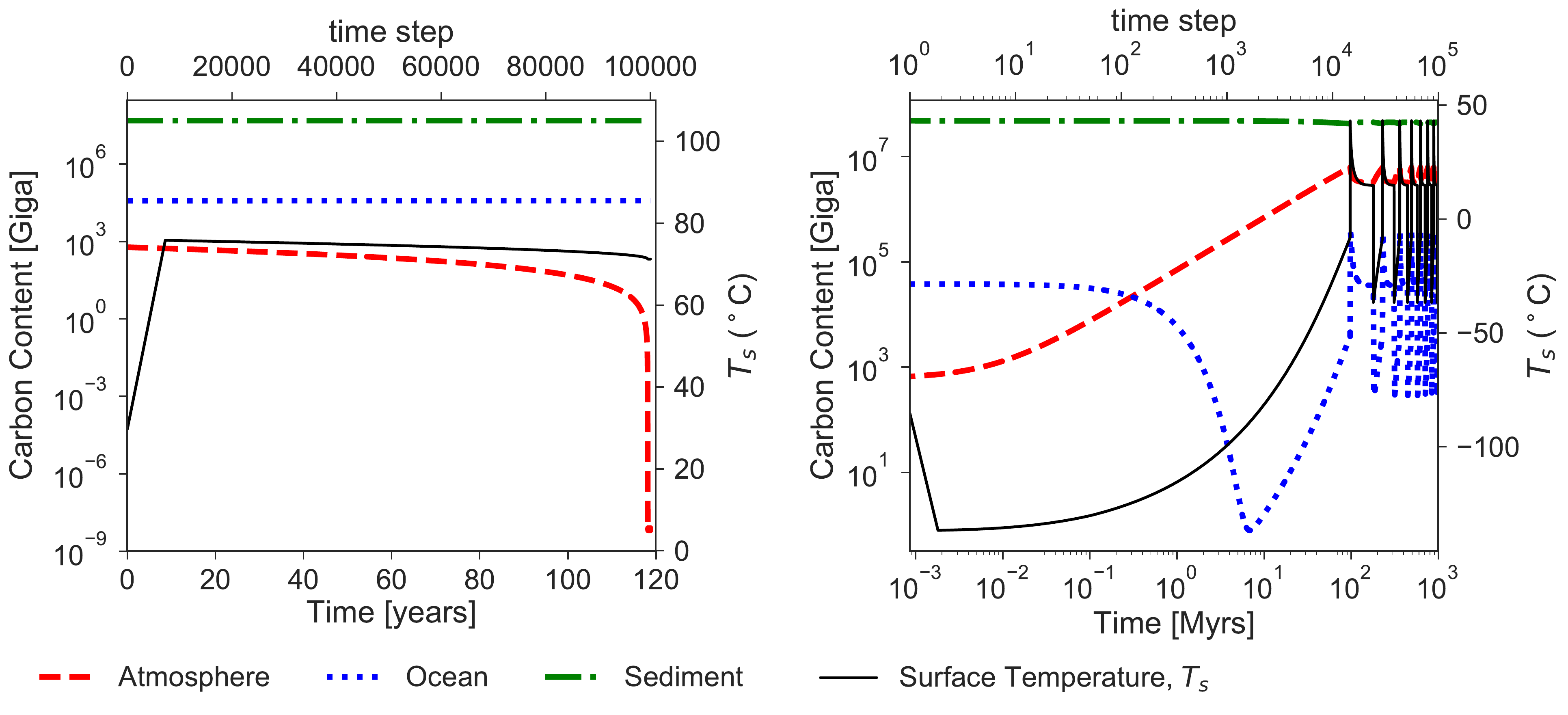}}
    \caption{Model evolution at the inner and outer edges of the habitable zone for a planet with land fraction $\gamma = 0.3$ and degassing $\Gamma = 1$. The inner edge (left) is defined as where the carbon reservoirs just reach equilibrium within the maximum model running time of 100,000 steps. This corresponds to a distance of 0.84\,au from the Sun. The outer edge (right) is where the equilibrium surface temperature oscillates around freezing and corresponds to a distance of 1.67\,au.}
    \label{fig:hzlimits}
\end{figure*}

The {\tt @earthlikeworld} twitter bot and classic toolkit on the {\tt Earth-Like} website allows users to vary the land fraction, $\gamma$, rate of degassing, $\Gamma$, and the position of the planet within the habitable zone, \emph{HZ}. The latter controls the equilibrium temperature of the planet in Equation~\ref{eq:temp}. An {\it advanced toolkit} is also available on the website that exchanges the \emph{HZ} parameter for the ability to select stellar type and the planet's distance from the star in astronomical units. This mode allows more experienced users the flexibility to investigate potential conditions on the Earth-sized planets we have discovered around different stars. However, the wider range of parameters allowed in the advanced toolkit risks the model failing to find an equilibrium surface temperature within the 100,000 steps allowed as a maximum running time on the website. Moreover, the model is normalised with respect to present day Earth values and therefore the results may become unrealististic for values far from those starting conditions. A warning is displayed on the advanced toolkit page to alert users to these issues, and the classic toolkit is recommended for most use. The parameter selection is listed in Table~\ref{table:parameters}. The interface is described in more detail in section `\nameref{sec:interface}'.

The website returns the new global surface temperature of the planet, an interactive plot displaying the evolution of the three carbon reservoirs and global surface temperature from present Earth values to the conditions on the new planet, and a visualisation of the planet (see section `\nameref{sec:visual}'). The plot is a visualisation of the model operation and allows users to see if the model has reached equilibrium and a table of the data can be downloaded within 5 minutes of creation (after this time, the file is deleted from the server). These results are supported by information outlining the carbon-silicate cycle, the parameters themselves and the model used.

When using the classic toolkit, the position of the planet can be selected within the  habitable zone. Of the planets we have discovered outside our Solar System, the worlds orbiting within the habitable zone have generated the widest interest due to their perceived potential for habitable conditions. Running the climate model for different locations within this region can highlight that surface temperatures can still be vastly different to what we experience on Earth today.

The edges of the habitable zone are defined for a planet orbiting a Sun-like star with the same land fraction as the Earth, $\gamma = 0.3$, and $\Gamma = 1$. For this climate model, the outer edge is the distance from the star where the global surface temperature drops below zero and the planet enters a freeze-thaw cycle. This occurs when the planet is at a distance of 1.67\,au from the Sun, in agreement with the outer edge of the habitable zone calculated using the more advanced climate models of \citet{Kasting1993} and \citet{hz}.

The freeze-thaw cycle is shown in the right-hand plot of Figure~\ref{fig:hzlimits}, which shows the model evolution for a planet placed at 0.999 of the habitable zone region (1.67\,au). The oscillatory behaviour occurs when the weathering rate is smaller than the degassing rate of ${\rm CO_2}$ back into the atmosphere when the global average temperature drops below freezing \citep{abbot2016} . If this situation is reversed (for example, by increasing the exposed land fraction, $\gamma$, and decreasing the degassing, $\Gamma$, in the model), then the planet goes into a snowball state with an average global temperature continuously below zero.

 \begin{figure}
    \centering
    \centerline{\includegraphics[width=0.8\columnwidth]{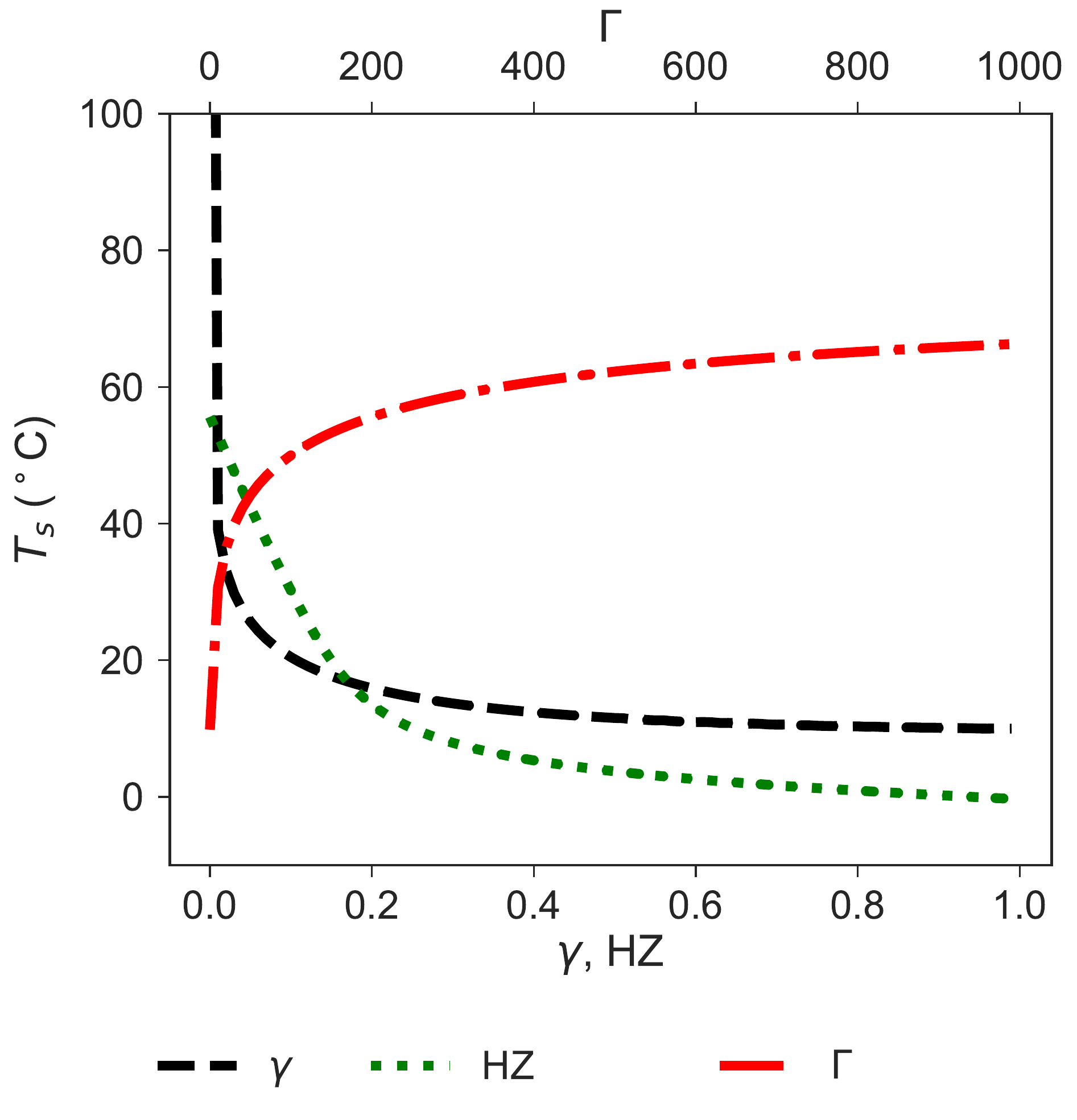}}
    \caption{The global surface temperature found when individually varying the land fraction, $\gamma$, position within the habitable zone region, \emph{HZ}, and rate of degassing, $\Gamma$.}
    \label{fig:parameters}
\end{figure}

During a freeze-thaw cycle, the planet initially cools as its equilibrium temperature has dropped from its original value at the present Earth's habitable zone fractional position of 0.19 (1\,au). The temperature dependence of the weathering, $F_{\rm w}$, results in the rate at which carbon is being drawn out of the atmosphere reservoir to decrease, allowing carbon to accumulate if the degassing rate exceeds this slowed weathering. The build-up of atmospheric carbon allows the surface temperature to increase steadily until the planet becomes warm enough that its albedo increases due to low ice coverage (equation~\ref{eq:albedo}). This results in a sharp rise in temperature as less radiation is reflected away from the planet, which is mirrored in the weathering rate and causes the abundance of carbon in the atmosphere drop. The surface temperature then decreases again in response to the lower abundance of atmospheric carbon, dropping back below freezing and causing the albedo to rise and lower the temperature still further. The weathering rate slows, carbon increases once again in the atmosphere and the process repeats.

 The inner edge of the habitable zone is defined as the closest location to the star where an equilibrium surface temperature can be found within 100,000 steps. The latter is a compromise between realistic behaviour and a practical time limit for running the model on a webserver. The temperature achieved at this inner edge is approximately $T_s \simeq 330$\,K, close to the moist-greenhouse (water-loss) limit in \citet{Kasting1993} and \citet{hz}, who find $T_{\rm moist} \sim 340 - 360$\,K as the point where water begins to be lost rapidly to the stratosphere. The inner edge for our model corresponds to a distance of $0.84$\,au around a Sun-like star. This is in reasonable agreement with \citet{Kasting1993}, who finds the moist-greenhouse limit occurs at $0.95$\,au but full runaway mode where oceans are lost from the planet entirely occurs at $0.84$\,au. The more recent models of \citet{hz} find limits of 0.99\,au and 0.97\,au for the moist-greenhouse and runaway-greenhouse respectively. When using the classic toolkit, the model therefore will show the carbon content in the atmosphere decreasing sharply and flattening out to a constant value in the last few steps, with a global temperature around $330$\,K ($54$\,$^\circ$C). This can be seen in the left-hand plot in Figure~\ref{fig:hzlimits}, where the planet has been placed at 0.001 of the habitable zone region (0.84\,au). The atmospheric carbon content flattens for the last few data points, which can be confirmed as a true equilibrium by running the model past the 100,000 step limit. With the advanced toolkit, the same conditions can be found by specifying a G-type dwarf star and distance of 0.84\,au. In hotter conditions with the advanced toolkit, the temperature continues to increase but does not reach equilibrium.

The global surface temperatures at the edges of the habitable zone can be surpassed by changing the other two planet parameters, the land fraction, $\gamma$, and the degassing rate, $\Gamma$. The allowed land fraction range is $0 < \gamma < 1$, where a global ocean and desert are excluded as it would be impossible to have a carbon-silicate cycle in those circumstance. The allowed range of the degassing factor is $0 < \Gamma < 1000$, where no volcanism is excluded for the same reason that the carbon-silicate cycle would not function, and the upper limit is imposed to prevent choices extremely far from the present day Earth values used to normalise the model. The resultant surface temperature found when the three possible parameters are varied individually is shown in Figure~\ref{fig:parameters}. For the parameters not being varied along each line, the value selected is the one for present day Earth with $\gamma = 0.3$, $\Gamma = 1$ and \emph{HZ} = 0.19. Colder and hotter values than those in Figure~\ref{fig:parameters} can be found when varying the parameters in combination. For example, a degassing factor of $\Gamma = 0.001$, land fraction of $\gamma = 0.999$ and habitable zone position of \emph{HZ} = 0.999 produces a snowball Earth with a global surface temperature of 152\,K ($-121^\circ$\,C) at the end of the simulation. Notably, without changes in the degassing or habitable zone position, the surface temperature is reasonably stable for land fractions down to less than 1\%. This agrees with the work of \citet{Abbot2012}, who also found that the carbon-silicate cycle could regulate surface temperature so long as a small amount of land was exposed for weathering to occur.

\section{Visualising the planet}
\label{sec:visual}

\begin{figure}
    \centerline{\includegraphics[width=0.9\columnwidth]{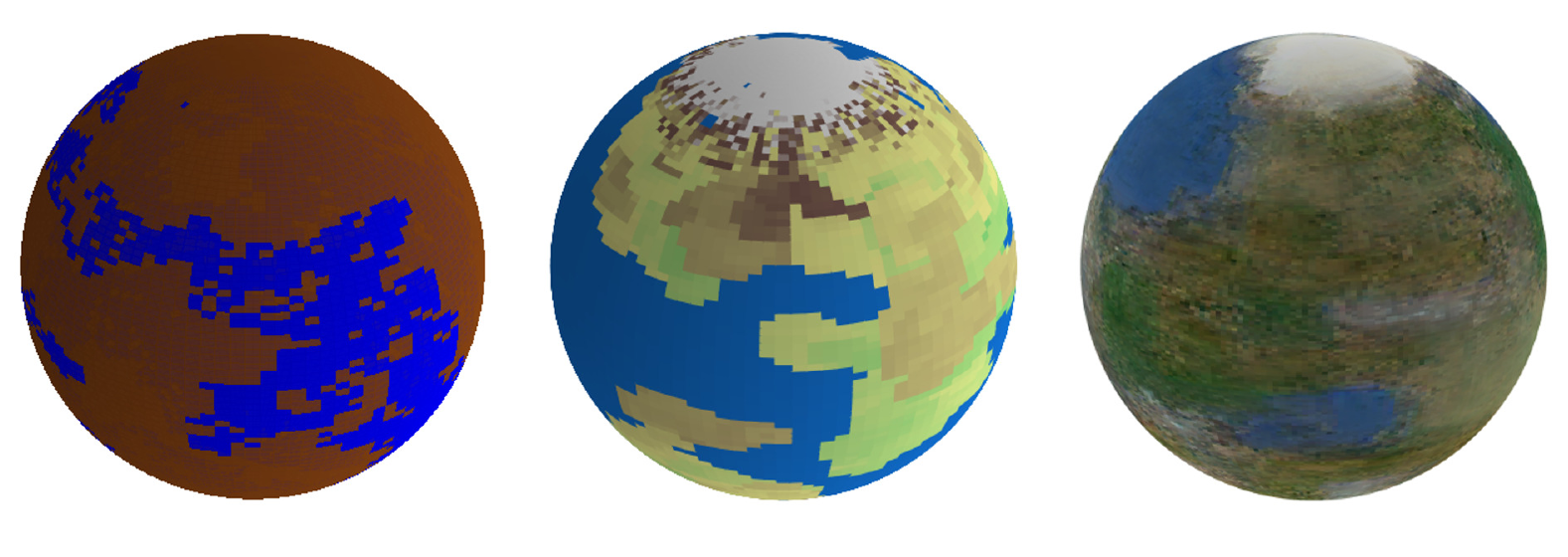}}
    \caption{Planet visualisations for a land fraction of $\gamma = 0.6$, degassing rate $\Gamma = 1$ and global surface temperature $T_s = 288$\,K ($15^\circ$\,C). The first version of the planet is shown on the left, where just the land fraction is represented in the ratio between brown and blue pixels. The middle and right-hand globe reflect the above three parameters, with surface temperature controlling the size of the ice caps and worlds with strong degassing having more yellow, brown and grey regions to represent the higher topology of a more volcanic landscape. The right-hand globe uses a neural network to create a more realistic image.}
    \label{fig:planets}
\end{figure}

\begin{figure*}[!t]
    \centerline{\includegraphics[width=0.9\textwidth]{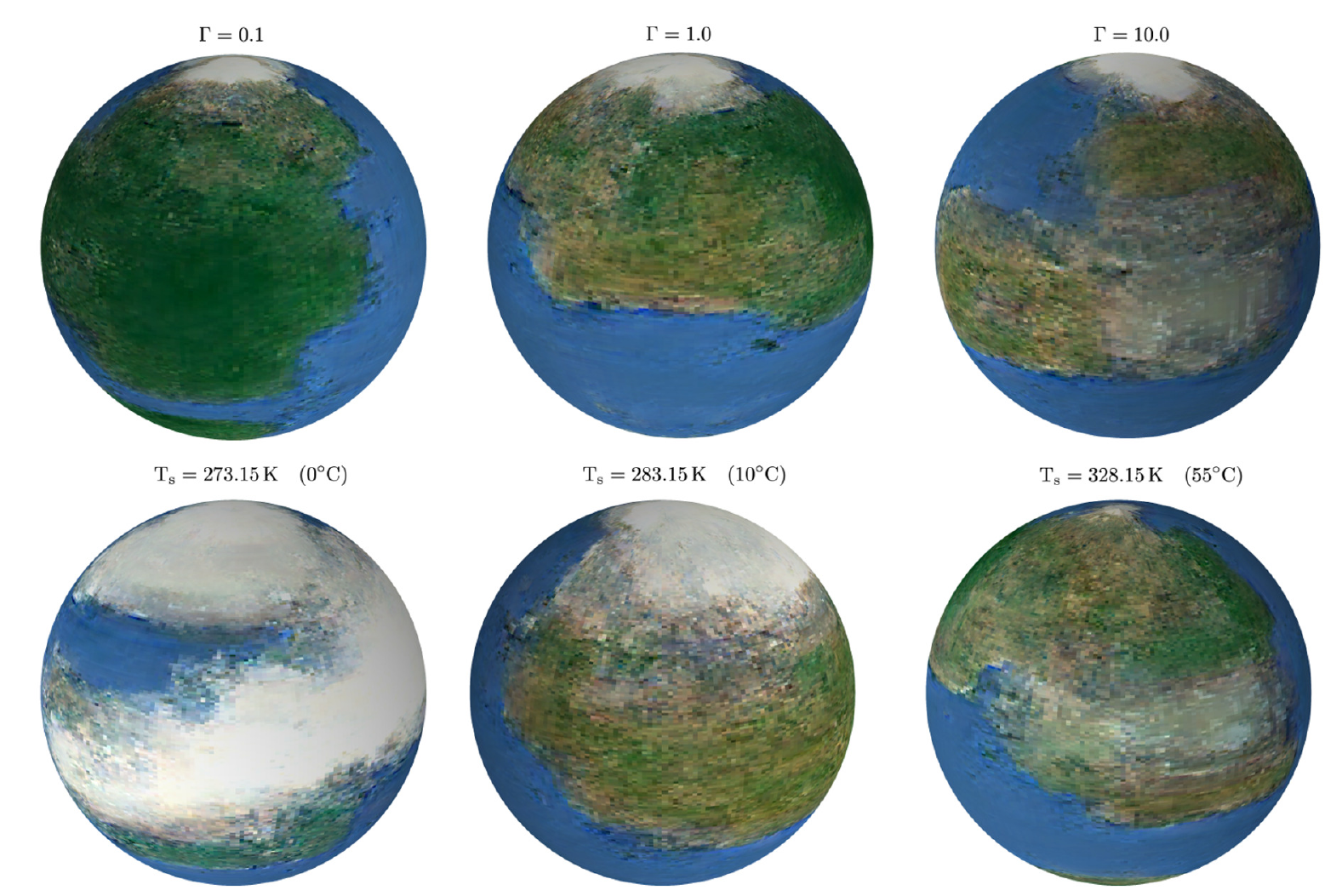}}
    \caption{Different visualisations of the {\tt Earth-Like} planet for varying properties of the degassing factor, $\Gamma$ (top row with land fraction $\gamma= 0.3$ and $T_s = 288$\,K ($15^\circ$C)), and global surface temperature (bottom row with $\gamma = 0.3$ and $\Gamma =1$).}
    \label{fig:planetsvaried}
\end{figure*}

Based on the planet parameters selected and calculated surface temperature, {\tt Earth-Like} creates a image of how the globe of the planet might appear. The true distribution of conditions on the planet will depend on many factors, such as the local incident radiation levels, which a box model is not able to estimate. The planet image is therefore a representation of the known parameters, designed to produce visual differences between the planets for the range of the parameter choices. The goal was to create a visual aid to increase engagement.

The planet visualisation was initially only produced when interacting with {\tt Earth-Like} through the twitter bot, \\${\tt @earthlikeworld}$. The earliest version was a static image of the globe, coloured in brown and blue to represent the land fraction, $\gamma$, of the planet. One example is shown in the left-most image in Figure~\ref{fig:planets}, for a planet with $\gamma = 0.6$.

An updated version is shown in the middle globe of Figure~\ref{fig:planets}, whose surface colours more fully represent the planet parameters. The surface map is initially divided into land and sea pixels, using a friends-of-friends scheme whereby pixels neighbouring land pixels also become land until the desired fraction $\gamma$ is reached. These are coloured by selecting a random value on a terrain colour map of greens, yellows, browns and grey, drawn from a Gaussian distribution. The Gaussian random number distribution is centred on a colour value based on the magnitude of the degassing factor, $\Gamma$. A more volcanic planet has an increased likelihood of selecting a brown or grey pixel for the land, compared with a green or yellow pixel. Additionally, neighbouring land pixels may not have steep gradients in colour in order to produce the more natural appearance of stretches of mountainous or valley regions across the map.

The surface temperature determines the size of an arctic zone, which stretches from the poles down towards the equator for values between 333\,K to 273\,K ($60^\circ$C - $0^\circ$C). At an average surface temperature of 273\,K, the entire planet is therefore within the arctic zone, whereas all the ice has melted by 333\,K and there is no arctic zone on the planet. Half of the arctic zone becomes a solid ice polar region for both land and sea pixels, while land within the arctic zone but at lower latitudes has a probability of selecting a white pixel that decreases with distance from the pole. For present-day Earth, the arctic zone is normalised to a latitude of $\pm 60^\circ$, with the polar (solid ice) region at $\pm 30^\circ$. The extent of this can be seen for the second two globes in Figure~\ref{fig:planets}.

An animated version of this globe that rotates to allow full view of the colour map was generated, with an axial tilt of $23.5^\circ$ in keeping with present day Earth. Based on the colours of the globe, it should be possible to make a ballpark guess at the parameters selected in the model.

The final visualisation of the planet was generated using a neural network to convert the pixelated globe into a realistic looking landscape. The neural network is a conditional Generative Adversarial Network (GAN) constructed along the lines of the pix2pix model \citep{Isola2016}. This type of network is composed of a {\it Generator} with a U-Net style architecture, and a {\it Discriminator} that uses a successive downsampling classifier architecture similar to VGG \citep{Ronneberger2015, Simonyan2014}. The job of the Generator is to take a reference image and return a transformed image, while the Discriminator takes both the reference image and transformed image as inputs and tries to determine whether the transformed image is the real image corresponding to that reference, or a fake image produced by the Generator. The Discriminator therefore provides supervision to the Generator to make its outputs more realistic. The loss function (which quantifies the robustness of the model) chosen was a Relativistic LSGAN, selected for its simplicity and stability \citep{Jolicoeur2018}.

The reference images used to create the {\tt Earth-Like} globe were satellite images of Earth terrain selected to include coastlines, mountainous and flat regions. The satellite images were downsampled by a factor of four to resemble the pixelated land maps produced based on the model parameters as described above. The Generator then tried to recreate the original image, while the Discriminator attempted to distinguish between the Generator image and the original satellite image. The satellite images were divided into a total 3250 patches of size 128 x 128 pixels and the network was trained on batches of 15. Once trained, the Generator was able to take a lower resolution version of the pixeleted middle globe in Figure~\ref{fig:planets} and return a land map like the right-hand globe in the same figure.

Because only the Generator is used in the final application, the Generator network needed to be relatively simple but the Discriminator could be made very large. One U-Net 'octave' was therefore used for the Generator (resulting in a total of five layers), but  five octaves were used for the Discriminator. In addition, because the target images being produced should be similar on average to the reference (as the reference is just a blurry vision of the target), the reference was added to the output of the last layer of the Generator to generate the transformed image (so if the Generator outputs 0, we recover the reference image). This accelerates training and reduces the amount of data needed.

Each planet visualised is unique, due to the stochastic creation of continents, highlands and icy regions. Planets with the same properties will share similar overall features, such as the size of the ice caps, but still have unique landscapes. Examples of the globes rendered for different parameter choices can be seen in Figure~\ref{fig:planetsvaried}. All six planets have a land fraction of $\gamma = 0.3$, with the top row showing variations in the degassing factor, $\Gamma$, for a surface temperature of 288\,K, while the bottom row has a constant $\Gamma =1$, but surface temperatures from freezing at 273.15\,K to a hot 328.15\,K. Increases in $\Gamma$ along the top row of planets from left to right shows the landscape changes from green to more yellow and then grey, suggesting volcanic mountain tops. Meanwhile, the ice recedes along the bottom row as the global surface temperature increases.

The visual appeal of the planet image generator made it a desirable addition to the website as well as the twitter bot. The main issue with the implementation was that image generation was slow, taking up to five minutes to complete; an impractical length of time for a website to load. The majority of the time was used in the rendering of the animation with the Python Matplotlib library. This was changed to use a static image of a sphere coloured according to the texture coordinates of a surface map. The image could then be modified with the surface map from the neural network for each frame of the animation, avoiding the sphere itself having to be re-rendered. This significantly reduced the time for animation production, but still produced an impractical overhead on the website if the user wanted to run multiple simulations in succession, such as in a classroom situation. This was tackled by making the creation of the visualised planet an optional addition after the website had loaded. Users can choose to run the planet image generator by pressing the "Render my planet" button on a static grey globe, which then launches the neural network in the background. Once complete, the blank globe updates to the rotating planet image.

\section{The {\tt Earth-Like} interface}
\label{sec:interface}

\begin{figure}
    \centerline{\includegraphics[width=0.9\columnwidth]{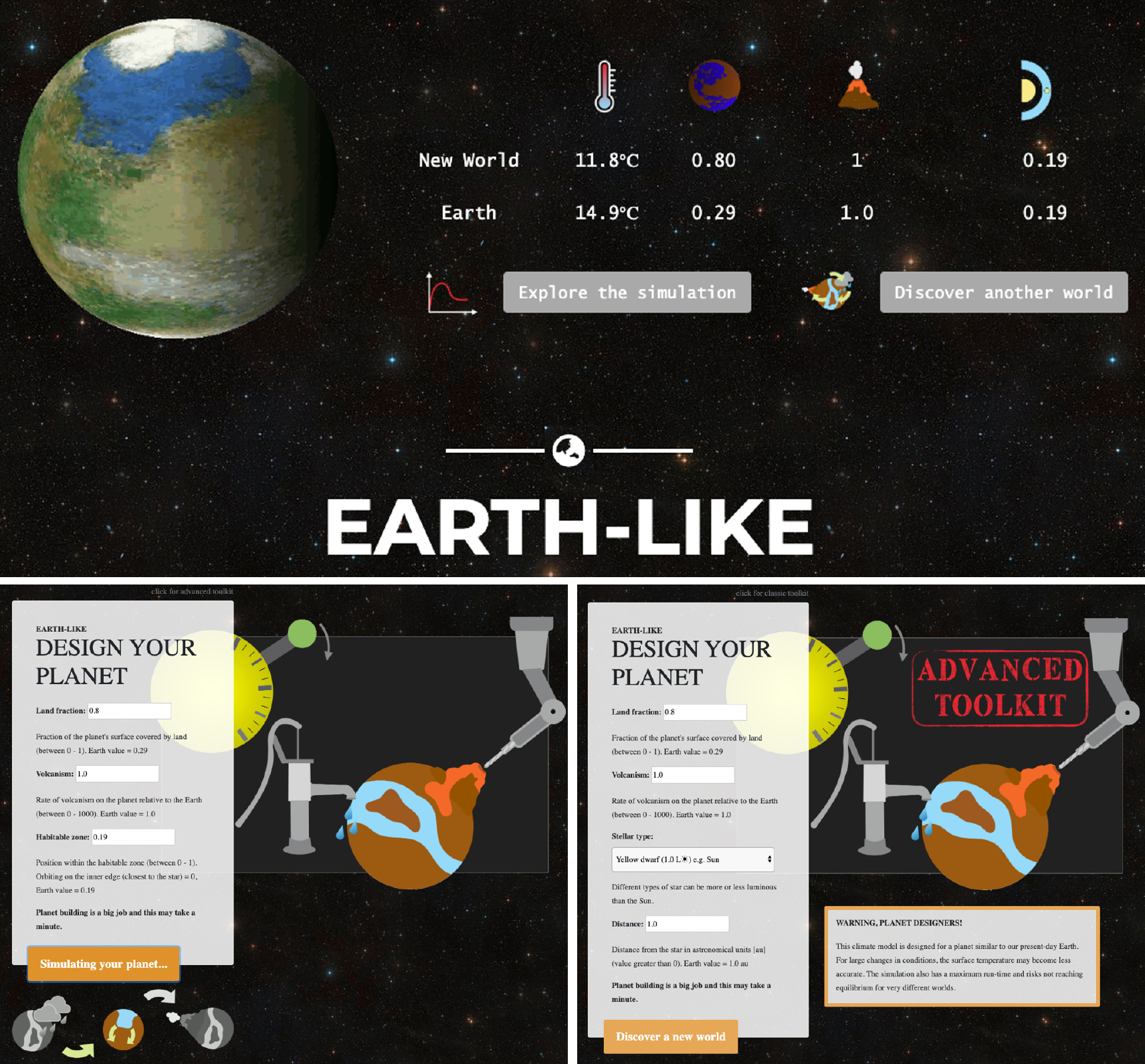}}
    \caption{The {\tt Earth-Like} website interface. Bottom two panels show the top page where the planet parameters can be chosen in the classic toolkit (left) and advanced toolkit. The top panel is the header of the main page, showing the calculated surface temperature and planet image. Below the header are the model details.}
    \label{fig:earthlikeinterface}
\end{figure}

The {\tt Earth-Like} model can be run on either the website, {\tt earthlike.world}, or through the twitter bot, {\tt @earthlikeworld}. Figure~\ref{fig:earthlikeinterface} shows the interface for the website; the classic toolkit is on the bottom-left panel, while the advanced toolkit is shown on the bottom-right. The top image in Figure~\ref{fig:earthlikeinterface} shows the header on the main page of the website once the model has run. The surface temperature of the planet is displayed, along with the properties that were selected, and the present-day Earth values for comparison. On the left of the header is an animated globe, created as described in section `\nameref{sec:visual}'.

Although the website speed was greatly increased by allowing the neural network to be run in the background after the page loaded, there is still a short delay due to needing run the climate model and create plots and images. Based on suggestions (see section `\nameref{sec:feedback}'), we added a splash bar that shows a small animation to indicate the page is still loading. This is shown at the bottom of the lower-left panel in Figure~\ref{fig:earthlikeinterface}. The form submit button text also changes to read `{\tt simulating your planet...}' after being pressed, to emphasise that a simulation is being run. This was also in response to initial feedback that suggested users were more engaged when they knew they were running a simulation, rather than accessing a finite set of pre-calculated options.

Below the header on the main page is information on the carbon-silicate cycle and climate model that includes plots similar to those in Figure~\ref{fig:hzlimits} for showing the model evolution. The section on the model is subdivided into a broad overview, the model evolution which includes the plots, and a mathematical description of how the carbon-silicate box model is solved. The aim was to appeal to a broad range of readers, all of whom might wish to understand the simulation but may have differing levels of mathematical training. In the overview, a diagram of the carbon-silicate cycle is shown that is similar to Figure~\ref{fig:boxmodel}. Arrows for the weathering and volcanism \& degassing change size and colour to reflect the chosen model parameters. For example, a location near the outer edge of the habitable zone or a large land fraction would result in a wide blue coloured arrow for the weathering flux, to indicate this would have a cooling influence on the planet.

\begin{figure}
    \centerline{\includegraphics[width=0.9\columnwidth]{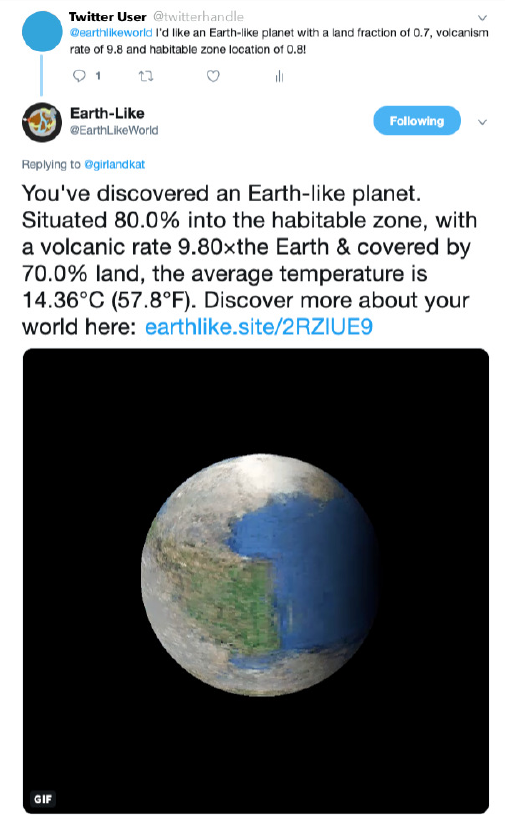}}
    \caption{The {\tt @EarthLikeWorld} interface on twitter. Parameters for land fraction ($\gamma$), volcanism (degassing factor, $\Gamma$) and habitable zone position (\emph{HZ}) are read by the twitter bot, which returns the Earth-like planet surface temperature, animated image and link to the main page of the website.}
    \label{fig:earthliketwitter}
\end{figure}

The twitter bot tweets a planet with randomly generated parameters several times a day. The tweet gives the values selected for the parameters for land fraction ($\gamma$), volcanism (degassing rate, $\Gamma$) and habitable zone position (\emph{HZ}) and includes an animated image of the planet globe created as in section `\nameref{sec:visual}'. Twitter users can also control the climate model themselves by tweeting at the bot with parameters for the land fraction, volcanism and habitable zone position. The bot searches for terms `{\tt land}', `{\tt volcan}' and `{\tt habitable}' and numerical values, associating values with parameters based on the order they appear in the text. For the land fraction and habitable zone position, the bot allows either a fraction or a percentage to be entered. Therefore, if you wanted to run the {\tt Earth-Like} model for a planet with land fraction $\gamma = 0.7$, degassing rate $\Gamma = 9.8$ and habitable zone location $0.8$, any of the following would be understood:

\begin{itemize}
    \item A land fraction of 0.7, volcanism rate of 9.8 and habitable zone location of 0.8.
    \item land fraction, volcanism, habitable zone of 0.7, 9.8, 0.8.
    \item 70\% land, volcano 9.8 and habitable position 80\%.
\end{itemize}

An example of a tweet and the reply is shown in Figure~\ref{fig:earthliketwitter}. If not all the parameters are specified, or the values chosen are out of range, then the bot selects a value at random for the missing parameter. If no parameters are specified (e.g. the tweet is `{\tt give me a planet!}') then the bot randomises all parameters and challenges the user to guess the values selected based on the image of the planet. All tweets provide a link to the main page of the website, which will load the same parameter values. The advanced tool kit cannot be run from the twitter bot.

\begin{figure}
    \centerline{\includegraphics[width=0.9\columnwidth]{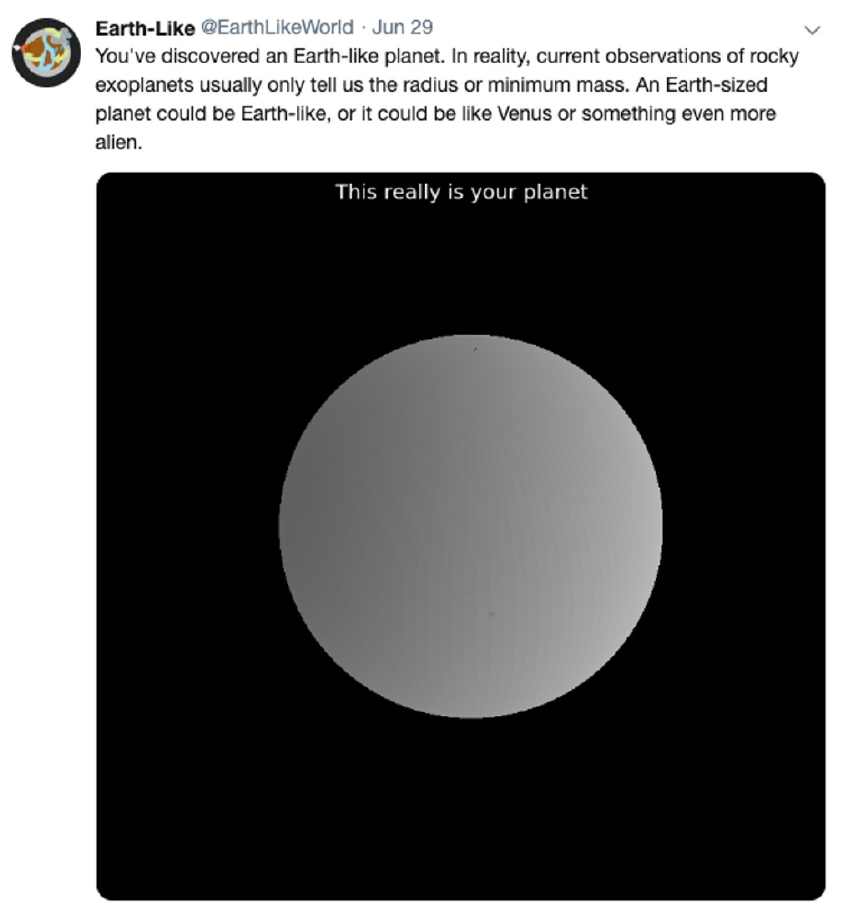}}
    \caption{There is a 1 in 20 chance of tweets from {\tt @EarthLikeWorld} returning a blank planet, reminding followers that our information about exoplanets is currently very limited. This occurs only when the twitter bot tweets independently, not in reply to a user tweet.}
    \label{fig:earthliketwitterblank}
\end{figure}

Since the twitter bot cannot contain information about the model, there was concern that the Earth-like planet images would suggest that all small planets are similar to our own; a message opposite to what the project wishes to portray. There is therefore a one-in-twenty chance that the twitter bot will return a simple grey image and note that current knowledge about exoplanets is limited to size, so we have no way of knowing if these worlds are truly Earth-like. This tweet is shown in Figure~\ref{fig:earthliketwitterblank}.

\section{Feedback}
\label{sec:feedback}

\begin{figure}
    \centerline{\includegraphics[width=0.9\columnwidth]{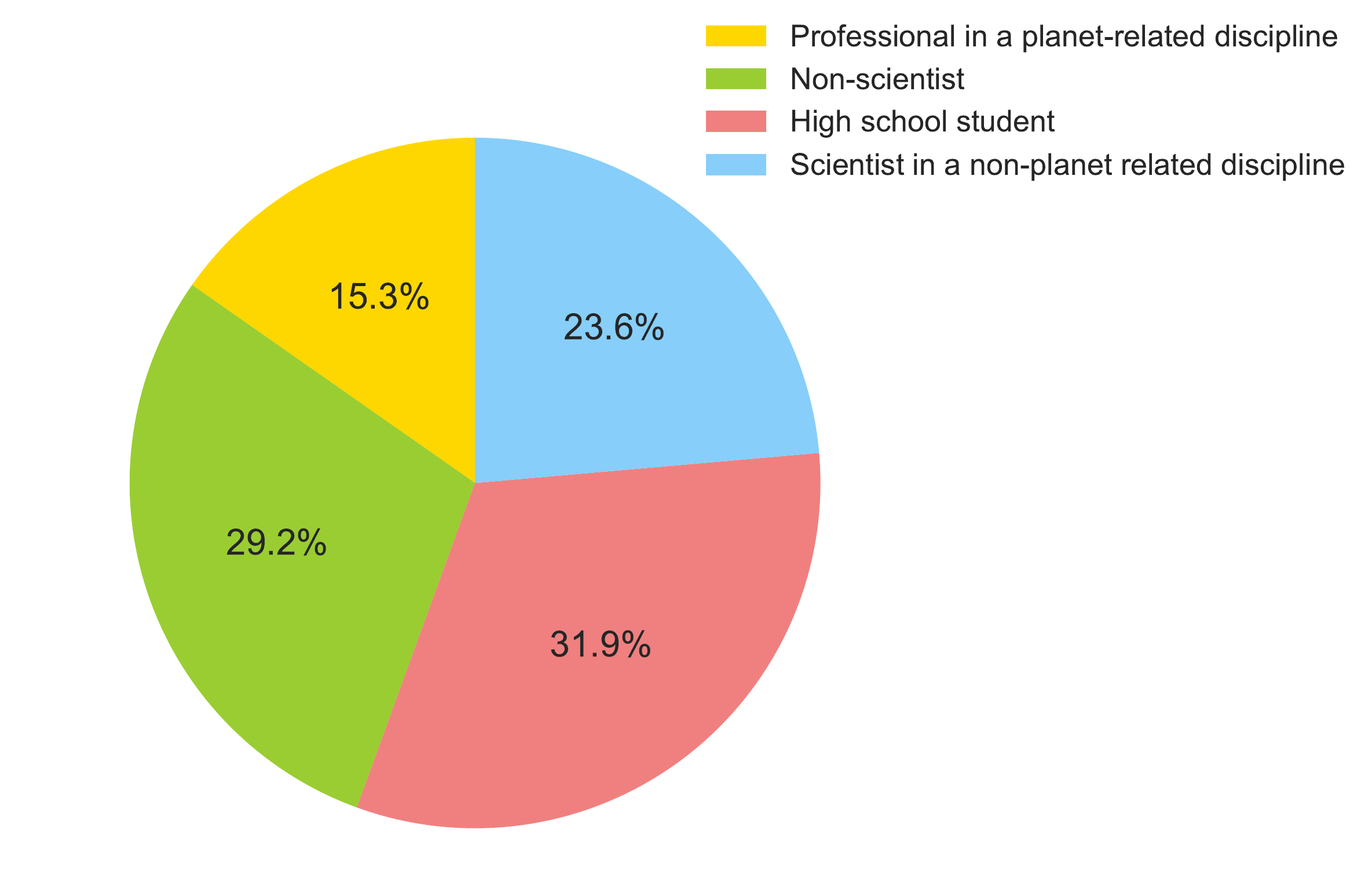}}
    \caption{Background of the participants who completed a short questionnaire on {\tt Earth-Like} from the website.}
    \label{fig:qdemographics}
\end{figure}

To gain preliminary feedback on the effectiveness of {\tt Earth-Like} in communicating information about planetary diversity, a questionnaire was attached to the website. 72 responses were received, including a class of US high school students and a US post-graduate class for in-service physics teachers. For both these classes, no extra information about the topic was provided apart from directing students to explore the website. The demographics for all participants is shown in Figure~\ref{fig:qdemographics}. About two-thirds of the participants can be considered not to have had scientific training beyond school. Just over half (53\%) also identified themselves as being an educator. The questionnaire posed four questions about the site and an opportunity to comment further at the end of the survey. The questions posed were:

\begin{enumerate}
    \item Which statement best describes your impression of\\ {\tt Earth-Like}? (5 possible answers)
    \item How did you find the information on the website? (4 possible answers)
    \item What did you think of the twitter-bot? (4 possible answers)
    \item You read in the media that `Earth 2.0' has been discovered: an Earth-sized planet orbiting another star in the habitable zone. What is the best description of this new world? (6 possible answers)
\end{enumerate}

\begin{figure*}
    \centerline{\includegraphics[width=0.9\textwidth]{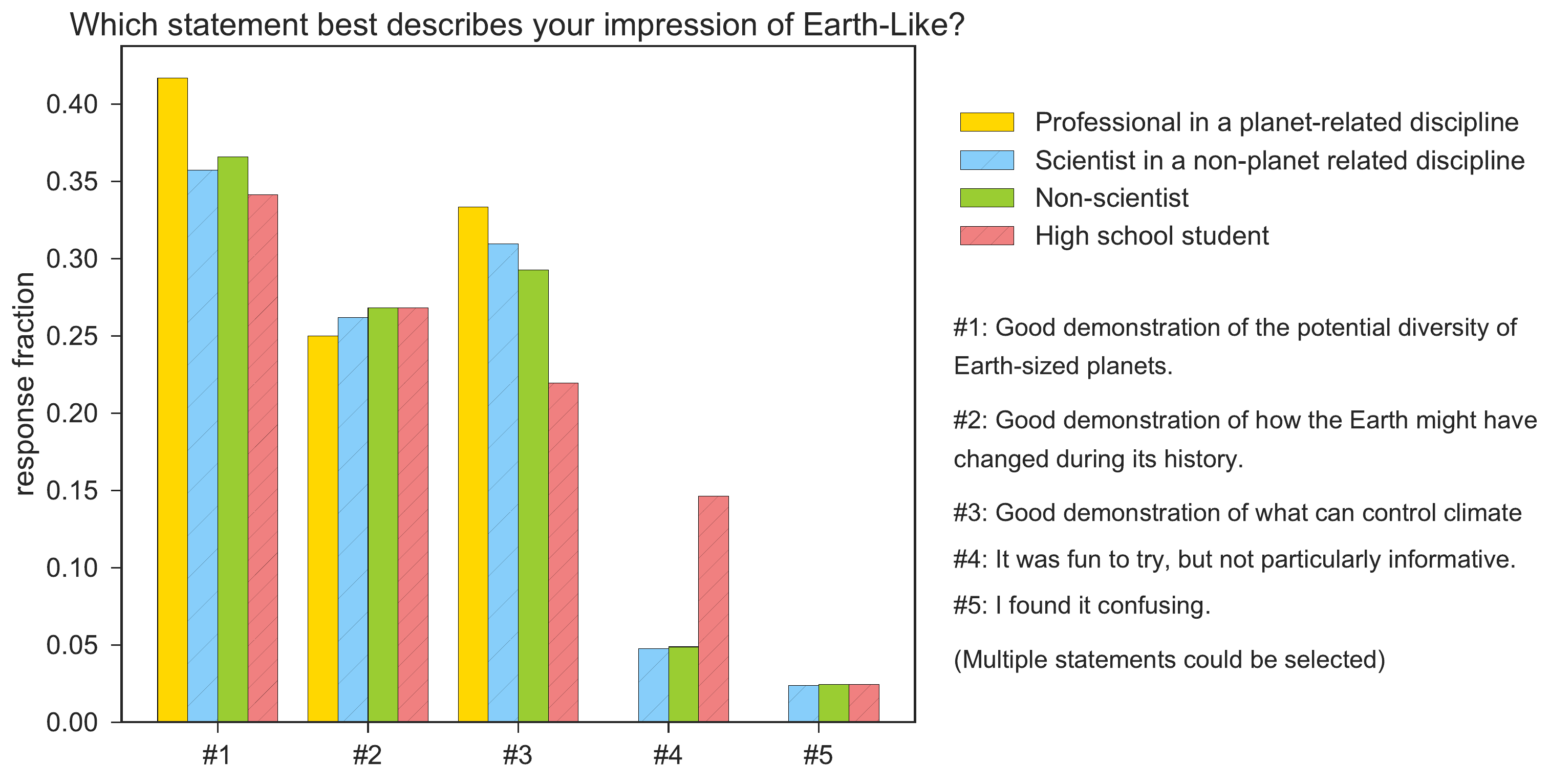}}
    \caption{Responses to question 1: `Which statement best describes your impression of {\tt Earth-Like}' on the website questionnaire, divided by demographic.}
    \label{fig:qstatement}
\end{figure*}

\begin{figure*}
    \centerline{\includegraphics[width=0.9\textwidth]{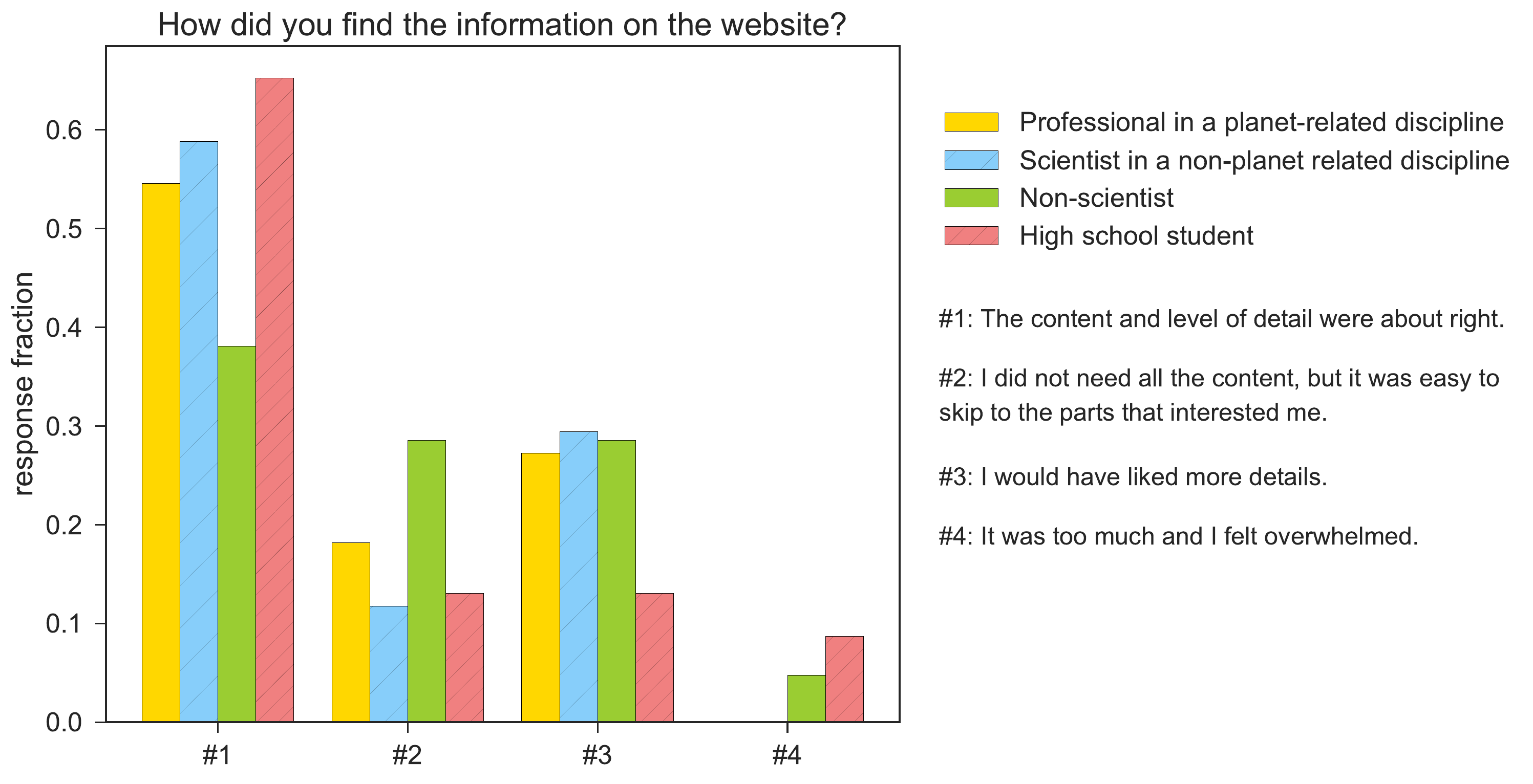}}
    \caption{Responses to question 2: `How did you find the information on the website?' on the website questionnaire, divided by demographic.}
    \label{fig:qinformation}
\end{figure*}

The first question targeted the overall message of the website, asking participants whether they felt the site had been informative by allowing one or more of five statements to be selected. These statements and the responses divided by demographic are shown in Figure~\ref{fig:qstatement}. The first three statements were consistent with the project goals, suggesting the site was a good demonstration of possible planet diversity, the evolution of the Earth and/or the factors that might affect a planet's climate. The last two statements deemed the site either uninformative or confusing. The majority of participants selected within the first three statements, suggesting the project is successful at transmitting information about planetary diversity. Just three people labelled the site confusing, although notably high school students struggled to understand the main message of the site, with six students saying the site was fun to use but not particularly informative.

Question 2 concerned the information available on the website, asking participants to select one of four possible statements that judged the available content to be too much or too little (Figure~\ref{fig:qinformation}). The majority of participants were happy with the level of detail, either finding the site sufficient or able to easily find the parts that focused their interest. A significant (24\%) of participants wanted more detailed information, which this paper will provide. This request is unsurprising for scientists who might potentially wish to use the site themselves for educational or outreach activities, but interestingly both the non-scientist and student groups showed a strong interest in having access to further information. Three participants from the non-scientist and student groups did find the site contained too much information that they could not easily filter.

\begin{figure*}
    \centerline{\includegraphics[width=0.9\textwidth]{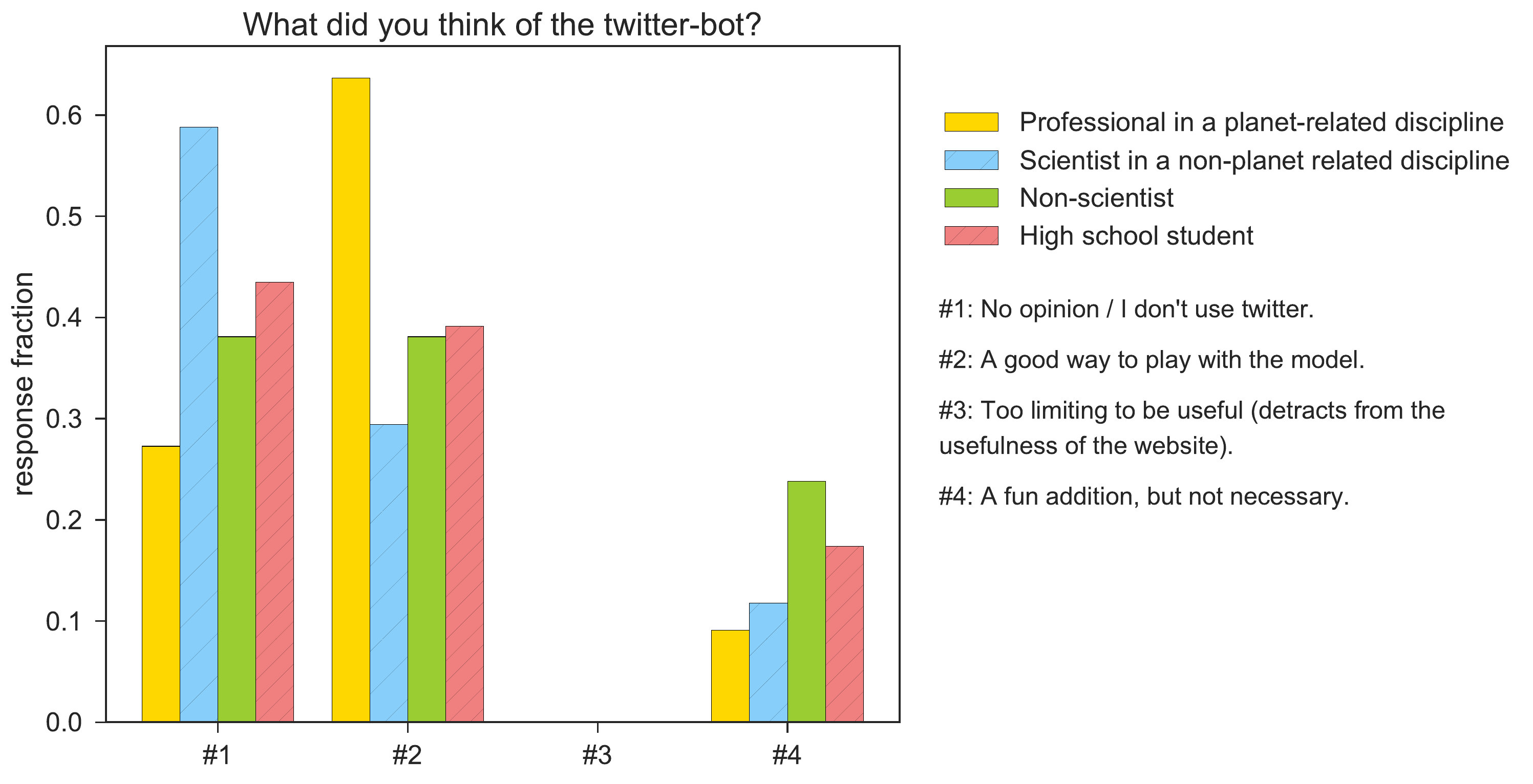}}
    \caption{Responses to question 3: `What did you think of the twitter-bot?' on the website questionnaire, divided by demographic.}
    \label{fig:qtwitter}
\end{figure*}

The third question shown in Figure~\ref{fig:qtwitter} asked about the twitter bot. Since the questionnaire was on the website, not all participants had used the twitter bot. Of the responses from people who had explored both the website and twitter bot, the twitter bot was predominately thought a good way to play with the model and was particularly popular among the category of professionals in planet formation disciplines. This could reflect that twitter is widely used in the scientific community for sharing information \citep{Holmberg2014, Ebner2009}.

\begin{figure*}
    \centerline{\includegraphics[width=0.9\textwidth]{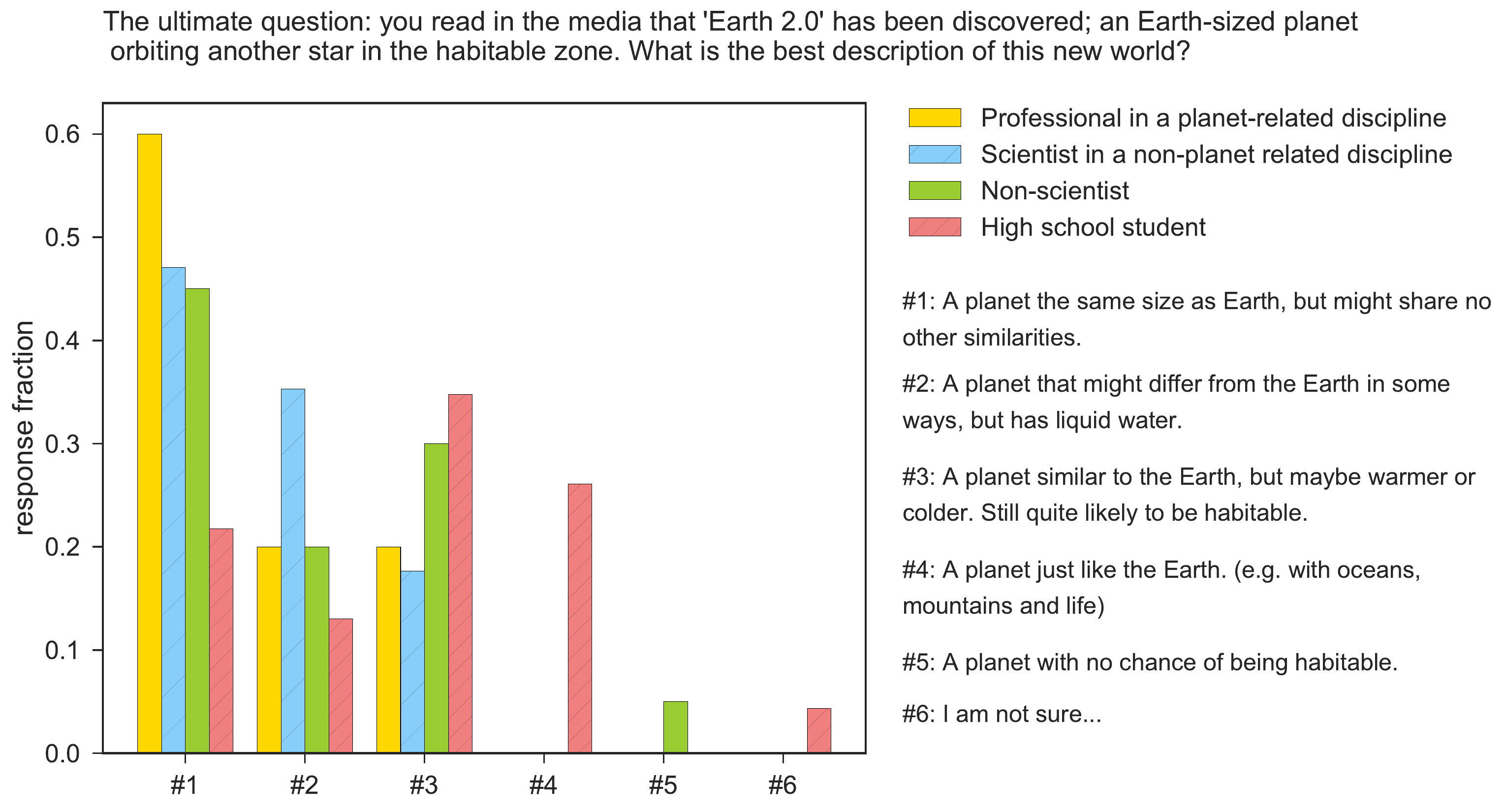}}
    \caption{Responses to question 4: `You read in the media that 'Earth 2.0' has been discovered: an Earth-sized planet orbiting another star in the habitable zone. What is the best description of this new world?' on the website questionnaire, divided by demographic.}
    \label{fig:qultimate}
\end{figure*}

The final question posed a test of the information learned on the {\tt Earth-Like} site. Media articles have frequently announced the discovery of an Earth-sized exoplanet that orbits in the habitable zone using terms such as {\it Earth 2.0}. Participants were asked which statement most accurately describes what is currently known about that planet. The statements and responses by demographic are shown in Figure~\ref{fig:qultimate}. The correct response was statement \#1: at present, we only know the radius or minimum mass of Earth-sized planets but nothing about their surface conditions. Therefore, all we can say for sure is that these planets have the same size as Earth but may not necessarily share any other similarities. Unsurprisingly, the majority of scientists selected the correct statement, as did participants in the non-scientist category. High school students did the poorest here, with the majority responding that the planet would be similar to the Earth and still likely to be habitable. This mistake may be due to assuming that the parameters presented by {\tt Earth-Like} represented all possible variables in planet diversity, as opposed to the demonstration of what a small subset of possible changes could produce.

\subsection{Response to feedback}

In response to the questionnaire results, two main additions were added to the website. The first was the creation of a chalkboard video. This five minute video describes what is currently known about exoplanets and the definition of the habitable zone. This information is also available in the text of the website, but it is possible that students in particular would find an animation a more enjoyable way to study. The video could also act as an introduction to a class, setting the scene for the framework of the website.

A {\it Frequently Asked Questions} was also added to the website based on comments from the questionnaire. Two common queries concerned why the phrase {\it Earth-like} was always printed below the planets produced by the model, even when surface temperatures were very different from present-day Earth, and what parameter choices would be suitable for Venus or Mars. The F.A.Q. answers points out that the changes to solar radiation level, degassing and exposed land fraction allowed by the {\tt Earth-Like} model are only minor changes to a planet, all of which the Earth has experienced during its history. All planets modelled are therefore extremely {\it Earth-like}, even if their surface temperatures are very different to current conditions on our planet. Conversely, neither Venus nor Mars have a carbon-silicate cycle, so they are not Earth-like enough to be represented by the model.

Responses to the questionnaire also mentioned that it was not initially clear an actual simulation was being run, rather than selecting planets from a list of pre-calculated possibilities. The suggestion was that emphasising this difference would encourage more experimentation. The form submission button on the front page was therefore changed to read `{\tt simulating your planet...}' when pressed. A splash animation was also added to indicate the model was running and the website had not crashed.

A request was made for a table of the simulation data to be available, so that classroom activities could be designed with students plotting different planet models alongside one another. As data could not be saved long term without requiring significant disc space on the server, a .csv file of the data was made available for five minutes after creation. Once this time had passed, the data is deleted.

While these additions hopefully assisted in clarifying the information on {\tt Earth-Like}, the responses from the high school student demographic suggest that the site would be most effective as part of a structured lesson where students where challenged to consider the ways a planet might differ from our own.

\section{Conclusions}
\label{sec:conclusions}

The search for a habitable planet is a goal that inspires people around the world, regardless of their scientific background. One of the challenges that the planetary and astrobiology community face is communicating the potential diversity of rocky planets, especially those that are frequently denoted {\it Earth-like} due to orbiting within the habitable zone. Successfully transmitting this information is essential for sharing current research progress in the field and thereby maintaining long term support from the public and government organisations, as well as encouraging young people to pursue STEM subjects.

The {\tt Earth-Like} website and twitter bot is an education and outreach tool for learning about planet diversity. The main goal is to promote understanding about what we currently know about Earth-sized planets and how different new worlds might be to our own.

The project presents this information using a simple interactive climate model that allows users to vary the land fraction, degassing rate and insolation level of an otherwise Earth-like planet with liquid surface water and calculate the resulting surface temperature. Running simulations is designed to be a more enjoyable and engaging way to discover that even small alterations to a planet's properties can have a major impact on the surface environment.

The model can be operated through both a website interface and a twitter bot. An {\it advanced tool kit} version allows users to select the stellar type and distance of the planet from the star for a greater range of possible values, with the understanding that this may significantly exceed the range of accurate results for the model.

The results are presented as a surface temperature value, graphical plot of the model evolution and an animation of how such a planet might appear. Details about the science are offered as a video, written description and F.A.Q.

Preliminary responses to the project suggest that the site is successful at providing information on planet diversity, although school classes might benefit from the website being part of a more structured lesson.

The {\tt Earth-Like} website is available at {\tt earthlike.world} and the twitter bot at {\tt @earthlikeworld}.

\ack[Acknowledgement]{The authors thank Colin Goldblatt, Joshua Tan and Dorian Abbot for their helpful advice and discussions. EJT was supported by JSPS Grant-in-Aid for Scientific Research Number 15K0514. JF was supported by JSPS KAKENHI Grant (16K05619).}

\bibliographystyle{apalike}
\bibliography{manuscript_arxiv.bib}

\end{document}